# BLIND CARRIER PHASE RECOVERY FOR GENERAL 2π/M-ROTATIONALLY SYMMETRIC CONSTELLATIONS

Emna Ben Slimane, Slaheddine Jarboui and Ammar Bouallègue

Laboratory of Communication Systems, National School of Engineers of Tunis, Tunisia
`emna.benslimane@yahoo.fr,slaheddine.jarboui@fss.rnu.tn,`
`ammar.bouallegue@enit.rnu.tn`

## ABSTRACT

*This paper introduces a novel blind carrier phase recovery estimator for general 2π/M-rotationally symmetric constellations. This estimation method is a generalization of the non-data-aided (NDA) nonlinear Phase Metric Method (PMM) estimator already designed for general quadrature amplitude constellations. This unbiased estimator is seen here as a fourth order PMM then generalized to $M^{th}$ order ($M^{th}$ PMM) in such manner that it covers general 2π/M-rotationally symmetric constellations such as PAM, QAM, PSK. Simulation results demonstrate the good performance of this $M^{th}$ PMM estimation algorithm against competitive blind phase estimators already published for various modulation systems of practical interest.*

## 1. INTRODUCTION

The need for non-data aided or blind feed forward carrier phase recovery in general 2π/M rotationally symmetric constellations systems is well established [1]-[2]. In order to satisfy this potential requirement, various estimation methods for L-ary QAM [3]−[6] and L-ary PSK [7]-[8] have been proposed in the literature. These blind estimators fit in either linear or nonlinear estimator group. The $M^{th}$ power-law estimator (PLE) [1] is a carrier phase estimator known as a maximum likelihood estimator at low SNR range. The PLE does not require any complex nonlinear optimizations but should have prior knowledge of the modulator constellation. Whereas, the well-known fourth-power estimator [2]-[9] is a special PLE designed for π/2 rotationally symmetric constellations such as QAM constellations. Furthermore, the minimum distance estimator (MDE) proposed by Rice & al [9]-[11] is considered as a straightforward nonlinear estimator that performs well with general QAM constellation at the cost of increased computational complexity. Recently, a blind carrier phase recovery estimator, referred to Phase Metric Method (PMM) has been originally proposed in [3] for fully modulated QAM transmissions. PMM is based on a special phase metric that exhibits an absolute minimum at the carrier phase offset. The performance of this algorithm achieves closely the Modified Cramér-Rao bound (MCRB) at both medium and high SNR range. Besides, this estimator requires fewer observed samples to come together with the MCRB by comparison to the aforementioned estimators.

The purpose of this work is to provide a generalization of this blind carrier phase estimator [3] for general 2π/M-rotationally symmetric constellations that encloses particular modulation systems of practical interest. As mentioned in reference [3], this NDA carrier phase recovery method was designed for both square- and cross-suppressed-carrier L-ary QAM constellations with quadrant symmetry [12]-[13]. Hence, the estimator presented in [3] is seen here as the fourth order PMM. In this work, we introduce the Phase Metric Method with $M^{th}$ order ($M^{th}$ PMM). This blind estimator is designed for general 2π/M rotationally symmetric constellations for full SNR range of practical interest. In order to evaluate the performance of this new estimator, we focus here on the fourth, eighth and sixteenth orders which can be applied to the QPSK, 8-PSK and 16-PSK modulation signals. These constellations are respectively π/2, π/4 and π/8 rotationally symmetric [14]. We study also the fourth order for the V.29 constellation which is π/2 rotationally symmetric. Simulation results demonstrate the efficiency of the PMM against the $M^{th}$ PLE and the MDE estimators.

The rest of this paper is organized as follows: in Section 2, the received signal model is presented. Then detailed description of the asymptotic performance of the $M^{th}$ PMM is depicted in Section 3 together with adequate computational-complexity reduction technique. The performance analysis of the novel method for π/2, π/4 and π/8 symmetry constellations is given in Section 4. Finally, Section 5 is devoted for the conclusions drawn from this work.

## 2. Discrete-time signal model

We consider a baseband frequency synchronized communication system over additive white Gaussian noise channel. The modulation interval $T$ is considered as perfectly known at the receiver side. Assuming a constant channel phase model, then any output sample of the modulation channel $r_k$ at time $kT$ can be written as follows:

$$r_k = s_k e^{j\theta_0} + \eta_k, \quad k = 0,1,....,N-1. \qquad (1)$$

Where $s_k$ is the complex symbols of 2π/M-rotationally symmetry constellation of a unit average energy transmitted at modulation time $kT$, $\theta_0$ stands for the unknown carrier phase and $\eta_k$ is the complex white Gaussian noise with variance $\sigma^2 = N_0/2$ along each dimension. N denotes the observation window size. The average signal-to-noise ratio (SNR) is defined as follows:

$$SNR = \frac{E\{\|s_k^2\|\}}{E\{\|\eta_k^2\|\}} = \frac{1}{2\sigma^2}. \qquad (2)$$

Where $E\{.\}$ denotes the expectation operator.

## 3. Asymptotic performance of the $M^{th}$ Phase Metric Estimator

### 3.1. $M^{th}$ PMM estimator and phase metric

A blind carrier phase recovery algorithm usually provides an estimate $\hat{\theta}_0$ for the unknown phase error $\theta_0$ without actually detecting the transmitted sample set $\{s_k\}$ but only from the received samples set $\{r_k\}$. Note that for 2π/M-rotationally symmetric constellation, the random phase offset $\theta_0$ is recovered within a modulo 2π/M phase ambiguity. For higher phase error values special coding is usually added [15]. Without loss of generality, we assume that the unknown phase offset $\theta_0$ lies in $[0, 2\pi/M)$.

In order to estimate $\theta_0$, we use the phase-metric $M(\theta)$ firstly introduced in [3] for QAM signals :

$$M(\theta) = \sum_{k=1}^{N} \min_{a \in C} \| r_k e^{-j\theta} - a \|^2 . \qquad (3)$$

Where $N$ denotes the number of observed samples $a$ runs through the $2\pi/M$ rotationally symmetric constellation $(C)$ and $\theta$ is an eligible phase within the investigation interval $[0, 2\pi/M)$. The detector picks the particular angle $\hat{\theta}_0$ within $[0, 2\pi/M)$ that minimizes the phase-metric. $N$ should be suitably chosen so that the observed samples set involves all channel signals with equal probability.

Theoretical analysis given in [3] demonstrates that in absence of noise the phase-metric $M(\theta)$ shows a unique minimum at $\theta = \theta_0$; which implies that the $M^{th}$ PMM estimator is unbiased. Computer simulations shown hereafter make obvious that this $M^{th}$ PMM estimator stands unbiased in presence of noise.

In order to measure the performance of the phase-metric (3), we consider a finite set of $n$ discrete phases $\{\theta_p = p \frac{2\pi}{M \times n} ; 0 \leq p \leq (n-1)\}$ uniformly distributed in the interval $[0, 2\pi/M)$. Then the absolute phase shift $\Delta \theta$ separating two consecutive discrete phases can be expressed as follows:

$$\Delta \theta = \theta_{p+1} - \theta_p = \frac{2\pi}{M \times n} . \qquad (4)$$

Note that for noiseless case the absolute estimate error is no longer zero but is uniformly distributed within $[0, \Delta\theta/2]$. Thus the standard deviation of the estimated phase is expressed as follows:

$$\text{StDev}(\hat{\theta}) = \left[ \frac{2}{\Delta \theta} \int_0^{\frac{\Delta \theta}{2}} x^2 dx \right]^{\frac{1}{2}} = \frac{\Delta \theta}{2\sqrt{3}} . \qquad (5)$$

Substituting (4) into (5), the expression of the standard deviation of the estimated phase becomes as follows:

$$\text{StDev}(\hat{\theta}) = \frac{\pi}{M \sqrt{3n}} . \qquad (6)$$

As can be seen from (6), the standard deviation of the estimated phase depends both on the number $n$ of discrete phases and the phase ambiguity of the $2\pi/M$-rotationally symmetric constellation. Thus, for a given $2\pi/M$-rotationally symmetric constellation, it is important to determine the minimum samples number $n_o$ that involves the convergence of the $M^{th}$ PMM estimator and also guarantees optimal performance. For high $SNR$ range, the appropriate value $n_o$ can be established with respect to the well-known Modified Cramér-Rao bound (MCRB), approximated to $(2N \times SNR)^{-1}$. Bounding expression (6) by the square root of the MCRB, leads to the following:

$$\frac{\Delta \theta}{2\sqrt{3}} = \frac{\pi}{M \sqrt{3n_0}} \leq \sqrt{\frac{1}{2N \times SNR}} . \qquad (7)$$

Thus, $n_0$ is the optimal integer that verifies the following condition:

$$n_0 = ceil(\frac{\pi}{M}\sqrt{\frac{2N \times SNR}{3}}). \qquad (8)$$

It appears from (8) that for a given $2\pi/M$ rotationally-symmetric constellation, $n_0$ depends both on the number of observed samples $(N)$ and the *SNR* level. Figure 1 shows that the phase metric $M(\theta)$ defined in (3) admits a unique minimum performed for the QPSK, V.29 ($M = 4$) and 8-PSK ($M = 8$) constellations, where the phase offset $\theta_0 = 30°$ and $SNR = 20dB$. N is set to 64. In addition the number of discrete phases satisfies equality (8). Notice that for the V.29 constellation, used in fax modems [1], the constellation signals with average symbol energy of 13.5 are given by:

$$A = \{\pm(1 + j), \pm(3 + 3j), \pm(-1 + j), \pm(-3 + 3j), \pm 3, \pm 5, \pm 3j, \pm 5j\}$$

As shown in Figure 1, the phase metric for the QPSK, 8-PSK and V.29 constellations admits a unique minimum equal to $\theta_0 = 30°$. Thus, we can conclude that the $M^{th}$ PMM is an unbiased estimator.

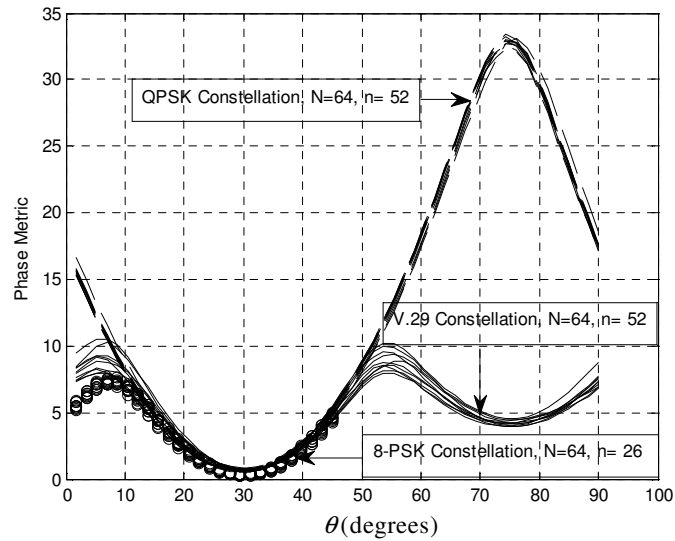

Figure 1. A phase metric for QPSK, 8-PSK and V.29 ($\theta_0 = 30°$, $SNR = 20dB$.)

As mentioned in the last paragraph, the discrete phases number satisfies expression (8) which shows that $n_0$ is small at low *SNR* levels then no meaningful complexity is involved. Whereas, at high SNR levels the optimal discrete phases number becomes larger and consequently the additive computational complexity increases.

### 3.2. Complexity reduction technique

In order to reduce the computational complexity of the $M^{th}$ PMM, we apply the same method described in [3]. In fact, we need to shorten the length of the investigation interval. The key has been found in [3] by using a multi-stage fourth PMM estimator for π/2 rotationally QAM constellation. We propose here a multi-stage $M^{th}$ PMM estimator for 2π/M rotationally-symmetric constellation as shown in Figure 2.

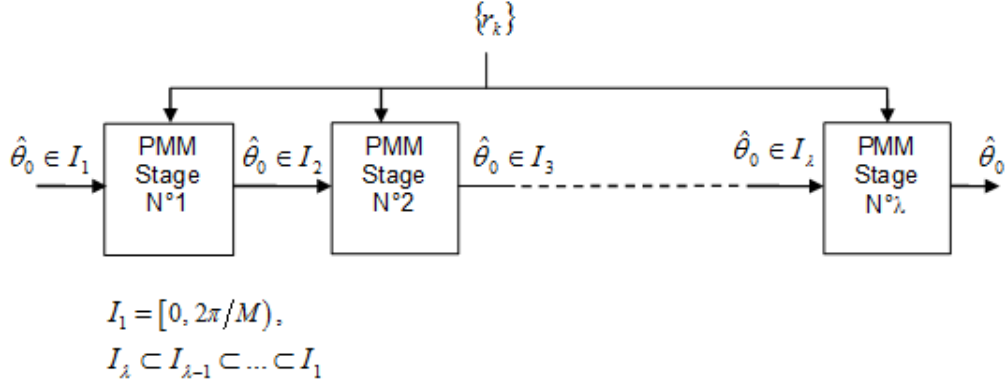

Figure 2. A multi-stage $M^{th}$ PMM estimator.

Each stage outperforms its forerunner, with tighter phase offset $\theta_0$ boundaries. For simplicity, let us consider the example of two-stage $M^{th}$ PMM estimator. The first stage estimates $\theta_0$ within the set $\{\theta_p\}$ as shown in (9) in which the discrete phases are assumed as uniformly distributed in $[0, 2\pi/M)$ :

$$\{\theta_p = p\frac{2\pi}{M \times n} \; ; 0 \leq p \leq (n-1)\} \tag{9}$$

Where $\theta_{\tilde{p}}$ $(0 \leq \tilde{p} \leq n-1)$ denotes the estimate of $\theta_0$ at the output of the first stage. Next, higher precision estimation of $\theta_0$ is pursued at the second stage by considering the subinterval $\left[(\tilde{p}-1)\frac{2\pi}{(M \times n)}, (\tilde{p}+1)\frac{2\pi}{(M \times n)}\right]$. Thus, the new set of discrete phases is uniformly distributed in the next subinterval given by:

$$\{\theta_q = (\tilde{p}-1)\frac{2\pi}{M \times n} + q\frac{4\pi}{M \times n^2} \; ; 0 \leq q \leq n\}. \tag{10}$$

With this procedure the optimal required number of discrete phases verifies the following equality:

$$n_0 = ceil(\sqrt{\frac{2\pi}{M}}\sqrt{\frac{2N \times SNR}{3}}). \tag{11}$$

Applying the same procedure as in [3] for general QAM constellations, we may reduce further the value of $n_0$ by considering higher number of PMM stages. In general, if we denote by $\lambda$ the number of PMM stages then for 2π/M rotationally-symmetric constellation, $n_0$ can be expressed as follows:

$$n_0 = Ceil\left\{\left(\frac{2^{\lambda-1}\pi}{M}\sqrt{\frac{2N \times SNR}{3}}\right)^{\frac{1}{\lambda}}\right\}. \tag{12}$$

According to [3], the computational cost of the $\lambda$-stage PMM for $2\pi/M$-rotationally symmetric constellation is given by:

$$\zeta_{\lambda-PMM} = \lambda \left( n_0 L (10N + 44) + 416 n_0 + 6 \right). \tag{13}$$

Next we investigate the optimal number of stages ($\lambda_{opt}$) that minimizes the computational cost $\zeta_{\lambda-PMM}$. Following the same method described in [3] we will solve this minimization problem graphically since $\lambda$ is integer. In order to evaluate the performances of the $M^{th}$ PMM for $2\pi/M$ rotationally-symmetric constellations, we consider the fourth, the eighth and the sixteenth orders PMM. For computer simulation purpose QPSK and V.29 constellations which are $\pi/2$-rotationally invariant are considered here for fourth PMM. For the eighth and the sixteenth orders, we consider respectively 8-PSK and 16-PSK constellations which are $\pi/4$ and $\pi/8$-rotationally invariant coded PSK, respectively.

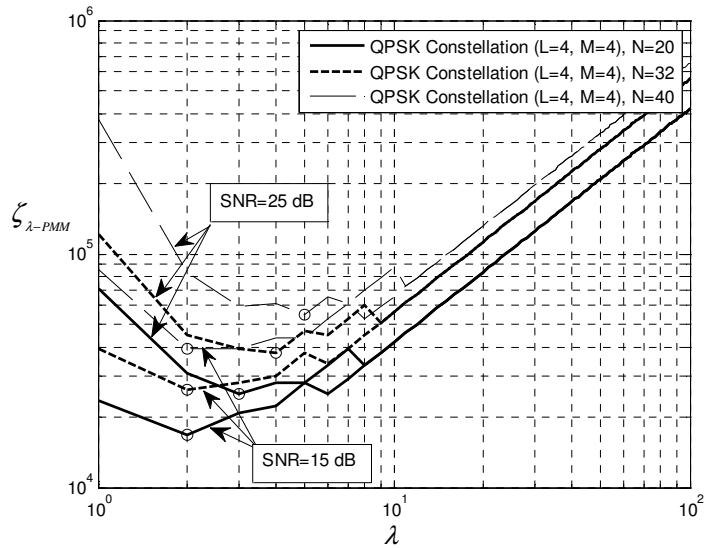

Figure 3. Computational latency against $\lambda$ for the QPSK constellation

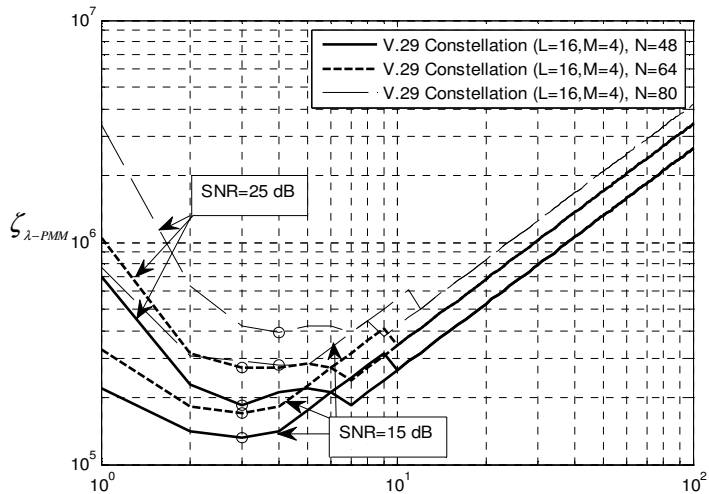

Figure 4. Computational latency against $\lambda$ for the V.29 constellation

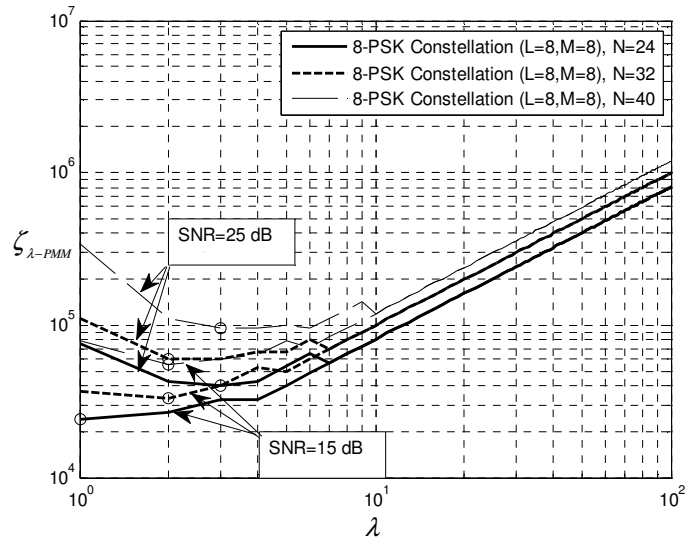

Figure 5. Computational latency against $\lambda$ for the 8-PSK constellation

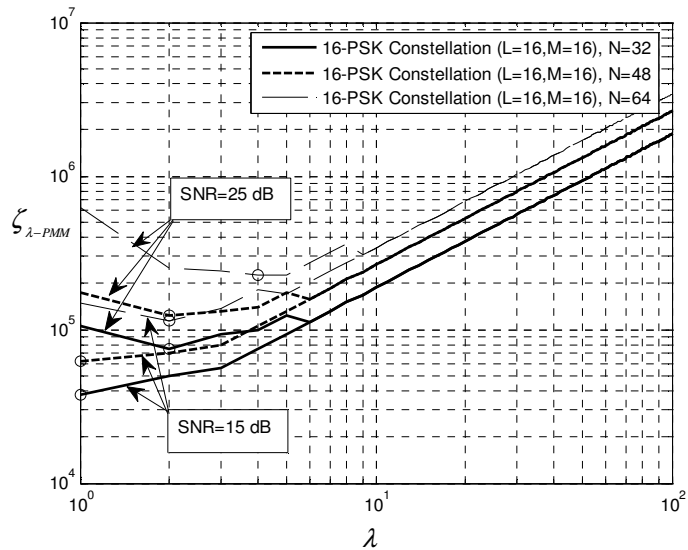

Figure 6. Computational latency against $\lambda$ for the 16-PSK constellation

The curves in Figures 3, 4, 5 and 6 reveal the computational cost $\zeta_{\lambda-PMM}$ against $\lambda$ for different SNR levels and different observed samples size $(N)$ with respect to QPSK, V.29, 8-PSK and 16-PSK constellations, respectively. The number of observed samples is chosen to be a multiple of the signal constellation size. The minimum of each curve is pointed by a circle. From these curves, we notice that the optimal number of M$^{th}$ PMM stages where $M = 4, M = 8$ and $M = 16$ for the considered constellations varies from one to five. In particular, $\lambda = 2$ guarantees acceptable computational complexity level for the practical SNR range [15 dB, 25 dB]. According to [3], for high order $\pi/2$-rotationally invariant coded QAMs (32-QAM, 64-QAM and 128-QAM constellations) the optimal stages number $\lambda_{opt}$ is four. Thus, we can conclude that for $2\pi/M$-rotationally symmetric constellation, multi-stage $M^{th}$ PMM can provide minimum computational complexity. Therefore, in the rest of this paper we consider a 2-stage $M^{th}$ PMM.

### 3.3. Choice of discrete phase number

Considering 2-stage PMM for $2\pi/M$ rotationally-symmetric constellation, the optimal discrete phase $n_0$ can be expressed as follows:

$$n_0 = Ceil\left\{\left(\frac{8\pi}{M}\sqrt{\frac{2N \times SNR}{3}}\right)^{\frac{1}{2}}\right\}. \qquad (14)$$

For a given constellation C, the optimal required number of discrete phases depends both on the SNR range and the number of observed samples N. If we fix the number N to at least four times the signal constellation size then $n_0$ becomes a nonlinear function of SNR. Yet, we restrict the study to the constellation symmetry $\pi/2$- for QPSK modulation signals where $M = L = 4$. Both Figures 7 and 8 provide simulation results of $n_0$ as function of both SNR and observed samples size $(N)$ for the QPSK modulation signals. But, in Figure 7, we consider a 1-Stage fourth PMM, and in Figure 8 the 2-Stage PMM is considered. Note that in Figure 8, we have not consider the ceiling function.

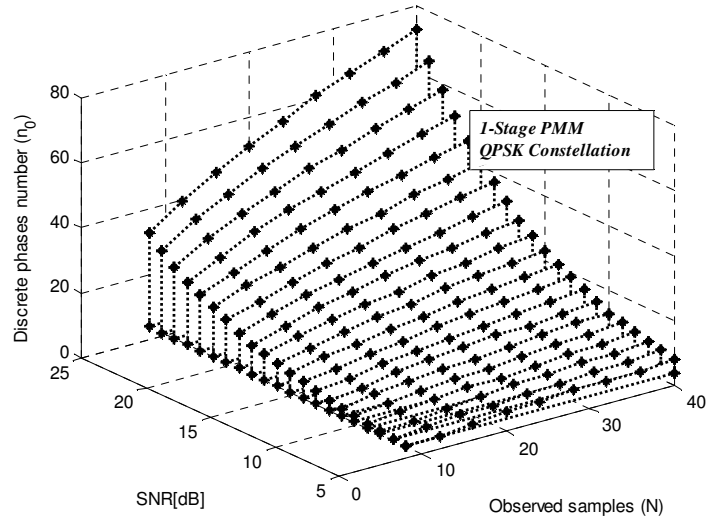

Figure 7. Minimum discrete phases $n_0$ against SNR for 1-Stage PMM (QPSK, M=L=4)

From Figures 7 and 8, we see clearly that for a given number of observed samples QPSK the discrete phases number increases as the SNR level grows. In such case, we choose the higher number of discrete phases corresponding to high SNR which remains valid for low SNR also. In addition, we remark that for a given SNR, the discrete phases number is a nonlinear increasing function of observed samples number N. Notice that the performance of the proposed estimator depend on the choice of $n_0$. Large number of phases can guarantee optimal performance but notably increases the computation time.

By comparing Figures 7 and 8 corresponding to one and two-Stage PMM respectively, we can conclude that the discrete phases number decreases to the 20% in the two-Stage case. This demonstrates that reduced number of discrete phases achieves optimal 2-Stage $M^{th}$ PMM performance.

Thus, as shown by Figure 8, the required phases number for the QPSK constellation that guarantees optimal performance is smaller than 12 for the SNR range between 5 and 25 dB and also for observed samples range between 8 and 40. If we choose 32 received QPSK signals, then we can see that ten discrete phases are sufficient to achieve optimal PMM performance for QPSK constellation. Notice that ten discrete phases remain a valid choice for the 8-PSK, the 16-PSK and the V.29 constellations. For the rest this work we set the discrete phases number to ten.

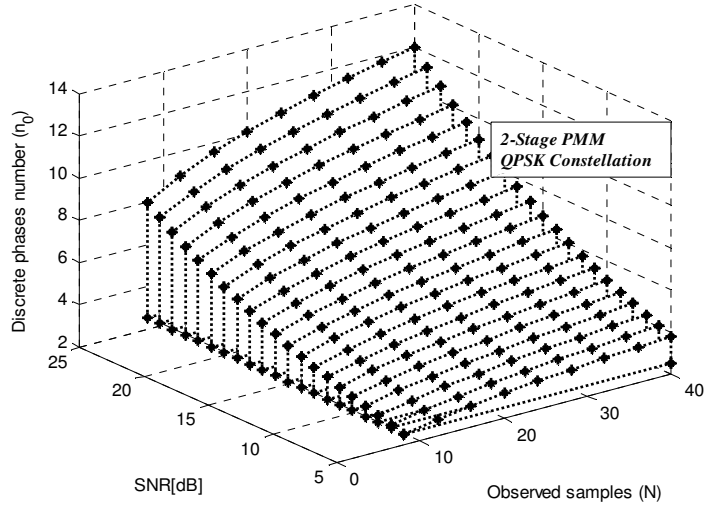

Figure 8. Minimum discrete phases $n_0$ against SNR for 2-Stage PMM (QPSK, M=L=4)

## 4. Performance comparison between $M^{th}$ PMM, MDE and PLE

In this section, we examine two competitive phase estimators for PSK constellations which are the minimum distance estimator [9] and the power law estimator [1]. The performance of the PMM estimator described in this work is analyzed both for QPSK, 8-PSK, 16-PSK and V.29 constellations and evaluated against the MDE and the PLE.

### 4.1. Power-Law Estimator

Designed for $2\pi/M$-rotationally symmetric constellations, the non-data-aided $M^{th}$ power-law phase estimator introduced by Moeneclaey and de Jonghe is known to be the maximum likelihood estimator as the signal to- noise ratio (SNR) goes to zero [1]. The PLE algorithm is known as monomial-based Viterbi and Viterbi synchronizer [16]. NDA feedforward carrier phase estimate is given by the following expression:

$$\hat{\theta} = \frac{1}{M} \arg \left( E\left[ \left(s_k^*\right)^M \right] \sum_{k=0}^{N-1} r_k^M \right). \tag{15}$$

Where N is the length of the observed data block, $s_k$ is the transmitted symbol and M=8 for 8-PSK and M=16 for 16-PSK. For QPSK and V.29 constellations, M is equal to four.

### 4.2. Minimum distance estimator

Under the assumption that the used constellation is $2\pi/M$-rotationally invariant coded PSKs, the blind minimum distance estimator makes a hard decision about the received signals. In fact, the received samples undergo a rotation by $\theta = (\theta_1, \theta_2, ..., \theta_n)$ in the range $[-\pi/M, \pi/M]$ to obtain a series of $n$ hypothesis sets. The set signals $\hat{a}_{ik}, i = 1, 2, ..., n$ are obtained by making

International Journal of Wireless & Mobile Networks (IJWMN) Vol. 4, No. 1, February 2012

hard decisions for each of the hypothesis sets. The Euclidean distance $D_i$ between the received signal $r_k$ and the i$^{th}$ hypothesis set signals is calculated as follow [4]:

$$D_i = \sum_{k=1}^{N} \left\| r_k e^{j\theta_i} - \hat{a}_{ik} \right\|^2. \tag{16}$$

The minimum Euclidian distance is denoted $D_\ell$, $D_\ell = \min(D_1, D_2, ..., D_I)$. The $l^{th}$ hypothesis set is used to calculate the residual phase offset given by :

$$\hat{\phi} = \tan^{-1}\left( \sum_{k=1}^{N} \frac{\operatorname{Im}\left( x_{\ell k}^{\dagger} \hat{a}_{\ell k}^{*} \right)}{\operatorname{Re}\left( x_{\ell k}^{\dagger} \hat{a}_{\ell k}^{*} \right)} \right). \tag{17}$$

Where $x_{\ell k}^{\dagger}$ the phase corrected signal given by:

$$x_{\ell k}^{\dagger} = x_{\ell k} e^{-j(\hat{\theta}_\ell - \frac{\pi}{4})}. \tag{18}$$

Thus the estimated phase offset of the received signal is:

$$\hat{\theta} = \hat{\theta}_\ell + \hat{\phi}. \tag{19}$$

Thus, the MDE estimator is a straightforward method designed to provide good performance for any constellation at the cost of increased computational complexity [9].

### 4.3. Simulation results

In order to evaluate the performance of the M$^{th}$ PMM estimator, we simulate its standard deviation for the QPSK, the 8-PSK, the 16-PSK and the V.29 constellations versus the SNR. The asymptotic performance of the phase metric method is evaluated against the MDE and the power-law estimators. Note that the PMM and the MDE estimators share the same metric; both of them use discrete phases. The differences between the two estimators are first, the MDE makes a hard decision about the received samples for each hypothetical phase and chooses only one possibility to compute the residual phase in a second soft decision stage. The PMM estimator is composed of a soft decision stage only. The proposed estimator is also compared to the well-known power-law estimator [2].

The M$^{th}$ PMM has been simulated for uncoded QPSK, 8-PSK, 16-PSK and V.29 signals. Then, each signal is multiplied by $e^{j\theta}$, where $\theta$ is the phase offset drawn from random uniform distribution in the interval $[0, 2\pi/M)$ at each trial, where M=4, M=8 and M=16 for the fourth, the eighth and the sixteenth order, respectively. Finally, the transmitting signal is embedded in additive white Gaussian noise.

Simulation results evaluate the phase estimate variance versus the SNR as depicted in Figures 9, 10, 11 and 12, along with the well-known Modified Cramer-Rao Bound. The investigation of the phase-metric minimum is performed by considering a two-stage PMM estimator ($\lambda=2$). Furthermore, the number of observed samples depends on the constellation density. The number N is chosen here four times the number of constellation signals for 16-PSK and V.29 constellations. For the QPSK and 8-PSK constellations, N is chosen equal to 32 and 40 respectively. The number of discrete phases is chosen equal to ten $(n_0 = 10)$ for the four cases. This choice of $n_0$ satisfies the equality (14) for the whole SNR range [0 dB, 25 dB].

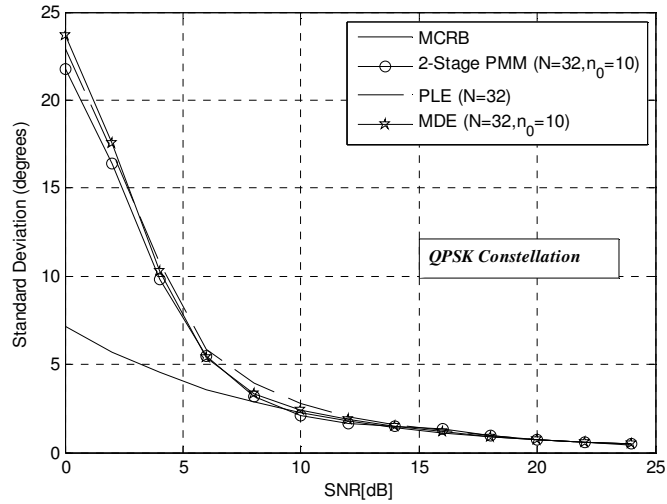

Figure 9. Phase estimate standard deviation for QPSK, N=32

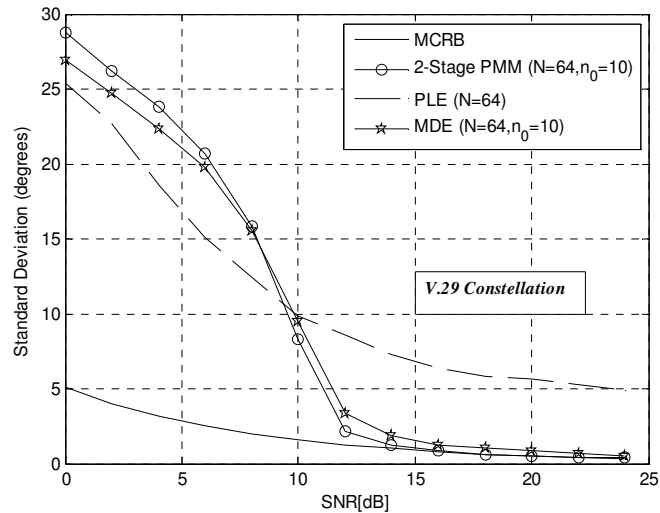

Figure 10. Phase estimate standard deviation for V.29, N=64

Figures 9, 10 and 11 refer to the QPSK, 8-PSK and 16-PSK constellations, respectively. These figures show that the PMM exhibits better performance than the MDE at low SNR for the PSK constellations. Figure 12 which refers to the V.29 constellation shows that the proposed method outperforms the MDE from SNR=8 dB. In addition, both the estimators approach the MCRB at high SNR levels. Notice that the proposed estimator offers much better flexibility than the MDE. As mentioned in [9], there is no clear criterion that can be adopted to fix the appropriate number of hypothetical phases in the first stage of the MDE, whereas for the proposed $M^{th}$ PMM estimator we use the well-known MCRB bound to determine the suitable number of discrete phases which makes the $M^{th}$ PMM estimator operate in a largely optimal manner.

The 2-Stage PMM is also compared to the PLE estimator. For the QPSK constellation, the fourth PMM exhibits substantially better performance than the fourth-power estimator. But, for the V.29 constellation, the PMM outperforms the fourth PLE from SNR=10 dB. Furthermore, for the 8-PSK and 16-PSK constellations, the eighth and the sixteenth PMM outperform the PLE estimator at eightth and sixteenth orders respectively.

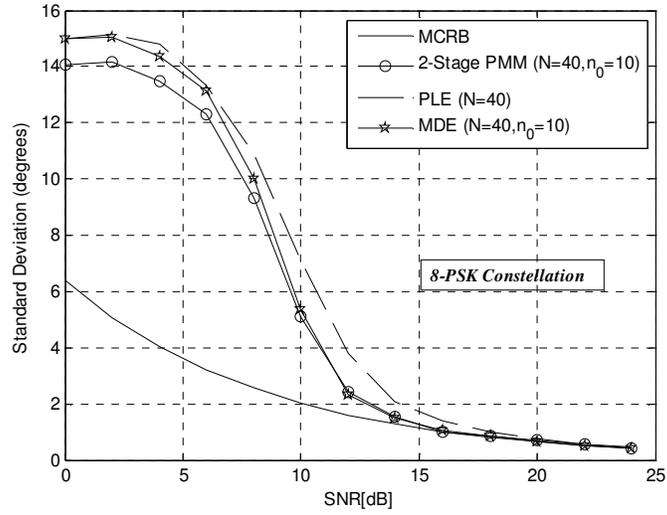

Figure 11. Phase estimate standard deviation for 8-PSK, N=40

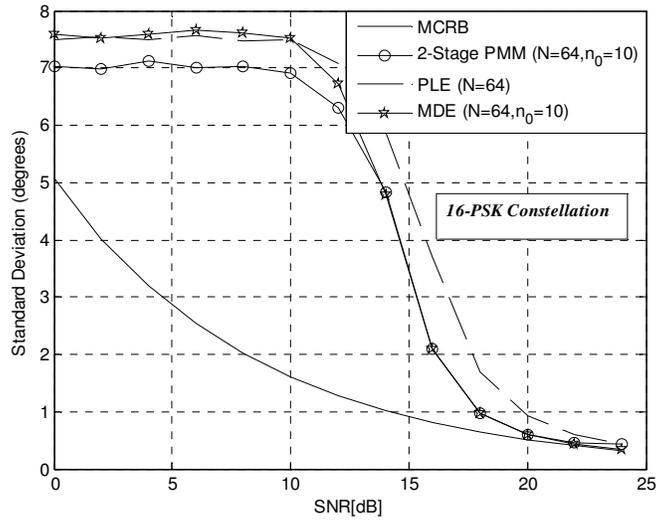

Figure 12. Phase estimate standard deviation for 16-PSK, N=64

## 5. CONCLUSION

In this contribution we propose a blind nonlinear phase recovery estimator for general 2π/M-rotationally symmetric constellations. This blind $M^{th}$ PMM estimator is a generalization of the fourth order PMM NDA estimation method [3] applicable to QAM constellations only. Theoretical results for general 2π/M-rotationally symmetric modulation system is presented. It is through the assessment of the performance of the fourth, the eighth and sixteenth PMM orders that our method is proven to be trustworthy. Simulation results also corroborate the theoretical performance analysis and indicate that the proposed optimal nonlinear estimator significantly outperforms the classic power-law and the nonlinear minimum distance estimators.